\documentclass[]{article}
\usepackage{amsmath}
\usepackage{graphicx}
\makeatletter
\@addtoreset{equation}{section}
\makeatother

\newcommand{\n}[1]{\label{#1}}
\newcommand{\eq}[1]{(\ref{#1})}
\newcommand{\eps}{\varepsilon}
\newcommand{\ba}{\begin{eqnarray}}
\newcommand{\ea}{\end{eqnarray}}

\def\ben{\begin{equation}}
\def\een{\end{equation}}
\def\bea{\begin{eqnarray}}
\def\eea{\end{eqnarray}}
\def\be{\begin{equation}}
\def\ee{\end{equation}}

\def\nn{\nonumber}

\input amssym.def
\input amssym.tex

\def\ft#1#2{{\textstyle{\frac{\scriptstyle #1}{\scriptstyle #2} } }}
\def\fft#1#2{{\frac{#1}{#2}}}
\def\cA{{\cal A}}
\def\bm{\bibitem}

\def\del{{\partial}}

\newcommand{\hoch}[1]{$\, ^{#1}$}

\newcommand{\auth}{\Large\bf{M. Cveti\v c\hoch{*}, G.W. Gibbons\hoch{\dagger},
D. Kubiz\v n\'ak\hoch{\dagger}
and C.N. Pope\hoch{\ddagger,\dagger}}}

\begin{document}

\begin{flushright}
\hfill {
DAMTP-2010-125\ \ \
MIFPA-10-56\ \ \
}\\
\end{flushright}

\begin{center}

{\Large{\bf Black Hole Enthalpy and an Entropy Inequality for the 
 Thermodynamic Volume}}

\vspace{25pt}
\auth

\large

\vspace{30pt}{\hoch{*}\it Department of Physics and Astronomy,\\
University of Pennsylvania, Philadelphia, PA 19104, USA}

\vspace{10pt}{\hoch{\dagger}\it DAMTP, Centre for Mathematical Sciences,\\
 Cambridge University, Wilberforce Road, Cambridge CB3 OWA, UK}

\vspace{10pt}{\hoch{\ddagger}\it George P. \& Cynthia W. Mitchell
Institute for\\ 
Fundamental Physics and Astronomy,\\ Texas A\&M University,
College Station, TX 77843-4242, USA}

\vspace{20pt}

\begin{abstract}

\rm

In a theory where the cosmological constant 
$\Lambda$ or the gauge coupling constant $g$ arises as 
the vacuum expectation value, its variation should be included in the first
law of thermodynamics for black holes. This becomes
$dE= TdS + \Omega_i dJ_i + \Phi_\alpha d Q_\alpha + \Theta d \Lambda$,
where $E$ is now the enthalpy of the spacetime, and  
$\Theta$, the thermodynamic conjugate of $\Lambda$, 
is proportional to an effective volume $V = -\frac{16 \pi
  \Theta}{D-2}$ ``inside the event horizon.'' Here we calculate
$\Theta$ and $V$ for a wide variety of $D$-dimensional
charged rotating asymptotically AdS black hole spacetimes, using
the first law or the Smarr relation.  We compare our expressions with
those obtained by
implementing a  suggestion of Kastor, Ray and Traschen, 
involving Komar integrals and Killing potentials, which we
construct from conformal Killing-Yano tensors. We conjecture 
that the volume $V$ and the horizon area $A$ satisfy the inequality
$R\equiv ((D-1)V/{\cal A}_{D-2})^{1/(D-1)}\, ({\cal A}_{D-2}/A)^{1/(D-2)}\ge1$,
where ${\cal A}_{D-2}$ is the volume of the unit $(D-2)$-sphere, 
and we show that this is obeyed for a wide variety of black holes, and
saturated for Schwarzschild-AdS.  
Intriguingly, this inequality is the ``inverse'' of the isoperimetric 
inequality for a volume $V$ in Euclidean $(D-1)$ space bounded by a
surface of area $A$, for which $R\le 1$.  Our conjectured {\it Reverse 
Isoperimetric Inequality} can be
interpreted as the statement that the entropy inside a horizon of a given
``volume'' $V$ is maximised for Schwarzschild-AdS. The thermodynamic 
definition of $V$ requires a cosmological
constant (or gauge coupling constant).  However, except in 7 dimensions, a
 smooth limit exists where
$\Lambda$ or $g$ goes to zero, providing a definition of $V$ even for
asymptotically-flat black holes.

\end{abstract}
\end{center}

\pagebreak

\tableofcontents

\section{Introduction}

 In theories where physical constants 
such as Yukawa couplings, gauge coupling constants  or Newton's constant 
$G$ and  the
the cosmological constant $\Lambda$  are not fixed {\it a priori},
but arise as vacuum expectation values and hence  can vary, their variation 
should be included in 
thermodynamic formulae such as the first law of black hole
thermodynamics. In fact such ``constants'' are typically 
to be thought of as the values at infinity of 
scalar fields. In the case of modulus fields,
the conjugate thermodynamic variables are scalar
charges \cite{GKK}.   The cosmological constant
$\Lambda$ behaves like a pressure,
\ben
P=-\frac{D-2}{16 \pi} \, \Lambda = \langle \, {\cal V}  \,   \rangle 
\,,\label{PLam}
\een    
where ${\cal V} $ is the potential of any scalars, and
the conjugate  thermodynamic variable $V$ is an effective volume
inside the horizon, or alternatively a regularised version of 
the difference in the total volume
of space with and without the black hole present.
\cite{BT1,BT2,CCK,KRT}.\footnote{The term $VdP$ is also  written as
  $\Theta d \Lambda$.}
Thus the first law of thermodynamics for black holes  reads   
\ben
d E= T dS + \sum_i \Omega _i d J_i + \sum_\alpha \Phi _\alpha  dQ_
\alpha  +V dP  
\label{law}
\een
and $E$ should be thought of as the total {\it gravitational  enthalpy},
which is the analogue of 
\ben
H=U + P V\,, 
\een
where $U$ is the total internal energy, so that
\ben
dU = T dS + \sum_i \Omega _i d J_i + \sum_\alpha  \Phi _\alpha
dQ_\alpha  - P  dV\,. 
\een
(Some further discussion of varying the cosmological constant in 
the black hole thermodynamical context has recently been given in 
\cite{lilulush,dolan}.)

  Of course if the cosmological constant is not treated as a variable,
then $H$, $U$ and $E$  coincide. 
However, even if the cosmological constant is not varied
the quantities $P$ and $\Theta$ enter the generalised 
{\it Smarr-Gibbs-Duhem
relation}, since $\Lambda$  affects the scaling properties  of the 
thermodynamic 
variables. The  Smarr-Gibbs-Duhem
relation is a simple consequence of the first law (\ref{law}), combined with
dimensional analysis. In $D$ spacetime dimensions  it reads \cite{CCK,GPP,KRT}  
\ben
E= (D-2)(TS+\sum _i \Omega_i  J_i)
 + \sum _\alpha \Phi_\alpha  Q_ \alpha  -\frac{2}{D-3} V P\,.  
\een
Moreover, in the simplest case of the Schwarzschild anti-de Sitter metric,
 and in  its single charged version, Reissner-Nordstr\"om
anti-de Sitter,  one finds that  
\ben\label{volume}
V  = \frac{A_{D-2}}{D-1}\, r_H^{D-1} \,,
\een
where ${\cal A}_{D-2}$ is the area of the unit $(D-2)$-sphere and 
$r_H$ is the radius of the horizon expressed in terms of the 
{\sl Schwarzschild radial coordinate}.
 
   These general considerations become especially 
interesting in the case of gauged supergravity and string theories, 
where the cosmological constant and the gauge 
 coupling constant $g$ are related by
\ben
\Lambda = -(D-1) g^2 \,,\qquad P = \frac{(D-2)(D-1)}{16 \pi}g^2 \,. 
\een 
Such theories can be obtained by means of sphere reductions
from the eleven or ten dimensional ungauged supergravity theories.
The most interesting gauged supergravities arise
in $D=4$,  obtained by an $S^7$ reduction from eleven dimensions;
in $D=5$, obtained by an $S^5$ reduction from ten dimensions;  and in $D=7$,
obtained by an $S^4$ reduction from eleven dimensions.
In these cases the cosmological constant and gauge coupling constant
are related to the curvature of the compactifying sphere; that is,
they are proportional to $({\rm radius})^{-2}$. 
Thus in these cases the term $V d P$ in the 
first law incorporates the thermodynamics
of the extra dimensional sphere, and its inclusion would be important
if the size of the extra dimensions, i.e. the radius of the sphere,
were to change with time.       

If one is contemplating time-dependent extra dimensions,
one should bear in mind that in descending from $(n+D)$ to $D$
spacetime dimensions on a compact manifold $K_n$ one has the relation 
\ben
G_D= \frac{G_{D+n}}{{\rm Vol}(K_n)}
\een
between the Newton constants.
Thus if $G_{D+n}$ is regarded as fundamental and hence unchanging, 
then if $ {\rm Vol}(K^n)$ changes with time, so will
$G_D$,  {\sl and its variation should also be contained in the first law}.

   In the remainder of this paper, we shall use the cosmological constant
$\Lambda$ rather than the pressure $P$ as the intensive thermodynamic
variable, and the conjugate extensive variable will be taken to be $\Theta$.
Thus the first law will be
\be
dE = T dS + \Omega_i dJ_i + \Phi_\alpha dQ_\alpha + \Theta d\Lambda\,,
\label{first}
\ee
where the metric in $D$ dimensions is asymptotically AdS, with the Ricci
tensor equal to (or, in the case of charged black holes, approaching)
$R_{\mu\nu} = \Lambda g_{\mu\nu}$.  We take $\Lambda=-(D-1) g^2$, where
for solutions in gauged supergravities, $g$ is the gauge coupling constant.
 
  From dimensional scaling arguments, the generalised Smarr relation is
\be
E = \fft{D-2}{D-3}\, (TS + \Omega_i J_i) + \Phi_\alpha Q_\alpha
   -\fft2{D-3}\, \Theta \Lambda\,.\label{smarr}
\ee
The pressure and cosmological constant are related by (\ref{PLam}), and
so $\Theta$ is related to the volume by
\be
\Theta = -\fft{(D-2)}{16\pi}\, V\,.\label{Thetav}
\ee

   In this paper, we investigate the role of the volume term in the
thermodynamics of asymptotically AdS black holes from various points of
view.  First of all, we note that since all the other quantities in
the generalised Smarr relation are already known, we can simply use
(\ref{smarr}) to furnish a {\it definition} of $\Theta$ in all the known
black hole examples.  This will necessarily also be consistent with the
generalised first law (\ref{first}).  It then becomes of interest to  
see whether $V$ calculated via (\ref{Thetav}) admits a natural physical
interpretation as a ``volume'' of the black hole.

   In simple cases such as a Schwarzschild-AdS or Reissner-Nordstr\"om-AdS,
it turns out that the volume calculated from (\ref{Thetav}) coincides
with a ``naive'' integration
\be
\int_{r_0}^{r_+} dr \int d\Omega \sqrt{-g}\label{nvol}
\ee
over the interior of the black hole, where the radial coordinate ranges
from the singularity at $r=r_0$ to the outer horizon at $r=r_+$.  In fact,
in such cases the volume $V$ turns out to be expressible as
\be
V= \fft{r_+\, A}{D-1}\,,\label{rAvol}
\ee
where $A$ is the area of the outer horizon.  With an appropriate modification
in the case that there are running scalar fields, a naive volume integration
again allows the potential $\Theta$ to be calculated for static charged
asymptotically AdS black holes.

  We find, however, that the situation becomes more complicated in the case
of rotating black holes.  If, for example, we consider the Kerr-AdS black
hole in $D$ dimensions, then a natural integration over the volume interior
to the horizon, of the form (\ref{nvol}), again, remarkably, gives rise
to the expression on the right-hand side of (\ref{rAvol}) (if one uses the
standard radial coordinate that appears in the metrics given in
\cite{gilupapo1,gilupapo2}).\footnote{In four dimensions, the 
notion of a ``black hole volume,'' 
obtained by integrating
$\int d^4x\sqrt{-g}$, was discussed previously in \cite{parikh,ballak}.}
However,
this volume, which we shall now call $V'$, is {\it not} the one that gives
rise to the correct thermodynamic potential $\Theta$.  Rather, it gives
\be
V' \equiv -\fft{16\pi}{(D-2)}\, \Theta' = 
  \fft{r_+\, A}{(D-1)}\,,
\ee
where $A$ is the area of the outer horizon.
$\Theta'$ is related to the true thermodynamic potential (defined via
(\ref{first}) or (\ref{smarr})) by
\be
\Theta'=\Theta + \fft1{2(D-1)}\, \sum_i a_i J_i\,,
\ee
where $a_i$ are the rotation parameters and $J_i$ the angular momenta of the
black hole.  We may refer to the associated volumes $V$ and $V'$ as the
``thermodynamic volume'' and the ``geometric volume'' respectively.

  Since in general we now have two different candidate definitions, 
it becomes of interest to investigate the
possible physical interpretations of each of the volumes $V$ and $V'$.
In view of the fact that the geometric volume for Kerr-AdS 
has the remarkable feature that $V'= r_H\, A/(D-1)$, as if it were just
the volume inside a sphere in Euclidean space, it is interesting to
test whether $V'$ and $A$ satisfy the Isoperimetric Inequality of
Euclidean bounded volumes.  Indeed, we find that 
\be
\Big(\fft{(D-1)V'}{{\cal A}_{D-2}}\Big)^{\ft1{D-1}}\le 
  \Big(\fft{A}{{\cal A}_{D-2}}\Big)^{\ft1{D-1}}\label{iso0}
\ee
for all Kerr-AdS black holes, with equality attained when the rotation
vanishes.  However, we find that for electrically charged black holes,
even without rotation (and hence $V$ and $V'$ are the same), the
isoperimetric inequality is violated.  

   If we instead use the thermodynamic volume $V$, then we find that
the isoperimetric inequality is {\it always} violated by rotating
Kerr-AdS black holes.  Furthermore, we find strong indications that
using $V$, the isoperimetric inequality is violated for all black
holes, with or without rotation and/or charge.  This leads us to 
conjecture that all black holes satisfy the {\it Reverse Isoperimetric
Inequality}, which asserts that
\be
\Big(\fft{(D-1)V}{{\cal A}_{D-2}}\Big)^{\ft1{D-1}}\ge 
  \Big(\fft{A}{{\cal A}_{D-2}}\Big)^{\ft1{D-1}}\label{iso1}
\ee
where $V$ is the thermodynamic volume of the black hole and $A$ is the
area of the outer horizon.  Equality is attained for Schwarzschild-AdS.

   The reverse isoperimetric inequality may be rephrased as the statement
that {\it for a black hole of given thermodynamic volume $V$, the
entropy is maximised for Schwarzschild-AdS}.

   The Smarr relation for black hole solutions of the vacuum Einstein
equations can be derived
by the Komar procedure, based on the integration of the identity 
$d{*d}\xi=0$ over a spacelike hypersurface intersecting the horizon, where
$\xi=\xi_\mu\, dx^\mu$ and $\xi^\mu$ is a Killing vector that is timelike
at infinity.  A generalisation to the case with a cosmological constant
$\Lambda$ has been discussed in \cite{bazzyl,Kastor,KRT}.  One writes $\xi$ in 
terms of
a 2-form Killing potential $\omega$ as $\xi={*d}{*\omega}$, and then
integrates the identity $d{*d}\xi + 2\Lambda d{*\omega}=0$ over the
spacelike hypersurface.  After using Stokes' theorem the integration
of ${*\omega}$ contributes a term on the sphere at infinity that removes a
divergent contribution from ${*d}\xi$ to give a finite expression for the
mass $E$, and a term on the horizon that furnishes an expression for $\Theta$.
One might hope that this could provide a further insight into the
question of whether the ``thermodynamic'' or the ``geometric'' $\Theta$ is
to be preferred.  Unfortunately, however, there is an ambiguity in the
definition of the Killing potential (the freedom to add a co-closed but
not co-exact 2-form to $\omega$), and this allows the expressions for
$E$ and for $\Theta$ to be adjusted in tandem.  As we discuss later,
the best that one can do is to choose a gauge for $\omega$ such that
the mass $E$ comes out to be the correct value, as already determined
by other means.  Necessarily, the integral yielding 
$\Theta$ then produces the ``thermodynamic'' expression rather than the 
geometric one. 

   We shall see that although the concept of the thermodynamic 
volume $V$ requires that one
consider an asymptotically AdS black hole in a theory with a
nonvanishing cosmological constant, it is
possible (except in $D=7$) to
take a smooth limit in the expression for $V$ in which the cosmological
constant is set to zero.  Since the thermodynamic volume still, in
general, differs from the geometric volume in this limit, one may
define, for an asymptotically flat black hole, the thermodynamic volume by 
first obtaining its expression in the more
general asymptotically AdS case, and then taking the limit where
the cosmological constant goes to zero. We find that this limit exists for
all the known asymptotically AdS black holes except for those in 
seven-dimensional gauged supergravity.  This case is exceptional because of
the existence of an odd-dimensional self-duality constraint in the
seven-dimensional theory.  It has the consequence that the volume
diverges in this case if the three rotation parameters and the electric charge
are all nonvanishing.

    The organisation of the paper is as follows.  In section 2, we 
use the Smarr relation, or, equivalently, the first law of thermodynamics,
to calculate $\Theta$ for the various static multi-charged black holes 
in four, five and seven dimensional gauged supergravities, and we show how
$\Theta$ is related to a volume integral of the scalar potential. In section
3, we use the same methods to calculate the thermodynamic expressions for
$\Theta$ for the rotating Kerr-AdS black holes in arbitrary dimensions.
We also show how these expressions are related to the geometric quantities
$\Theta'$ that are directly given by volume integrations.  We also
perform similar calculations for some examples of charged rotating 
black holes in four and five dimensional gauged supergravities.  In
section 4 we examine the isoperimetric inequality, and we show in
particular that the Reverse Isoperimetric Inequality holds for all the
black hole examples we have considered. In section 5 we review the 
derivation for the Smarr relation using the generalisation of the Komar
procedure, and then we give a detailed construction of the required
Killing potentials $\omega$ for Kerr-AdS, making use of the
conformal Killing-Yano tensors that exist in these backgrounds. The paper
ends with conclusions in section 6.  In an appendix, we present some
explicit results for the Killing potentials in four and five
dimensional Kerr-AdS.

\section{Static Charged Black Holes}

   In this section, we consider charged static black hole solutions in
gauged supergravities in $D=4$, 5 and 7 dimensions. We shall work in
conventions where Newton's constant is set to 1, and the action takes the
form
\be
I =  \int \sqrt{-g} \Big[ \fft1{16\pi} \,R -\fft1{16\pi}\, 
f(\phi) F^{\mu\nu} F_{\mu\nu}
- {\cal V(\phi)} +\cdots\Big]\,,
\ee
where $F_{\mu\nu}$ is a $U(1)$ field strength (there may be just one, or 
several), $f(\phi)$ represents the coupling of scalar fields, and
${\cal V(\phi)}$ is the potential term for the scalar fields.  In the 
solutions we shall consider, the scalar fields go to zero at infinity,
and then $f\phi)$ approaches 1,  and
the potential approaches
\be
{\cal V}\longrightarrow -\fft{(D-1)(D-2)}{16\pi}\, g^2\,,
\ee
where $g$ is the gauge coupling constant.
Thus the black hole solutions are asymptotic to AdS$_D$ with $R_{\mu\nu}
\longrightarrow -(D-1) g^2\, g_{\mu\nu}$.  Details of the black hole 
solutions can be found in \cite{Duff,BCS,10author}, where they were 
constructed, and further discussion of their thermodynamics can be
found in \cite{gibperpop2}.   In what follows, we summarise the
pertinent properties of the black holes for each of the dimensions
4, 5 and 7, and we calculate the quantity $\Theta$ in each case.

\subsection{Charged AdS black holes in $D=4$}\label{4chargesec}

   The metric, electromagnetic potentials, and scalars fields are 
given by \cite{Duff}
\bea
ds_4^2 &=& - \prod_{i=1}^4 H_i^{-1/2}\, f dt^2 + \prod_{i=1}^4 H_i^{1/2}\,
        \bigg(f^{-1} {dr^2} + 
        r^2 d\Omega_{2}^2 \bigg) \,,\nn\\
A^i &=& \fft{\sqrt{q_i(q_i+\mu)}}{2(r+q_i)}\, dt\,,\qquad
X_i = H_i^{-1}\, \prod_{j=1}^4 H_j^{1/2}\,,
\eea
where
\be
f = 1 - {2m \over r} + g^2 r^2 \prod_{i=1}^4H_i \,, 
\qquad H_i=1+{{q_i}\over r}\,.
\label{f4}
\ee
The four scalar fields $X_i$, subject to the constraint 
$\prod_{i=1}^4 X_i=1$, have the potential
\be
{\cal V}=-\fft{g^2}{16\pi} \sum_{i<j} X_iX_j\,.
\ee

   The relevant thermodynamic quantities are given by
\bea
E &=& m  +\ft14 \sum_i q_i\,,\qquad Q_i= \ft12 \sqrt{q_i(q_i+2m)}\,,\qquad
 S= \pi\prod_i (r_+ +q_i)^{1/2}\,,\nn\\
T &=& \fft{f'(r_+)}{4\pi}\, \prod_i H_i^{-1/2}(r_+)\,,\qquad \Phi_i =
   \fft{\sqrt{q_i(q_i+2m)}}{2(r_++q_i)}\,,
\eea
where the outer horizon is located at $r=r_+$, the largest root of $f(r_+)=0$.

   Substituting into the first law (\ref{first}) or the Smarr relation
(\ref{smarr}), we find that $\Theta$ is given by
\be
\Theta= -\fft{r_+^3}{24}\, \prod_i H_i(r_+)\, \sum_j \fft1{H_j(r_+)}\,.
\ee
With the integral of the scalar potential defined by
\be
W= \int_{r_0}^{r_+} dr \int d\Omega_2 {\cal V}\sqrt{-g}\,,
\ee
where $r_0$ is taken to be the largest root of
\be
4r_0^3 + 3 r_0^2 \sum_i q_i +2 r_0\sum_{i<j} q_i q_j +
   \sum_{i<j<k} q_i q_j q_k =0\,,
\ee
we find that $\Theta$ can be written as 
\be
\Theta = -\fft1{\Lambda}\, W\,.
\ee

\subsection{Charged AdS black holes in $D=5$}

   The metric, electromagnetic potentials, and scalars fields are
given by \cite{BCS}
\bea
ds_5^2 &=& - \prod_{i=1}^3 H_i^{-2/3}\, f dt^2 + \prod_{i=1}^3 H_i^{1/3}\,
        \bigg(f^{-1} {dr^2} +
        r^2 d\Omega_{3}^2 \bigg) \,,\nn\\
A^i &=& \fft{\sqrt{q_i(q_i+ 2m)}}{(r^2 +q_i}\, dt\,,\qquad
X_i = H_i^{-1}\, \prod_{j=1}^3 H_j^{1/3}\,,
\eea
where
\be
f = 1 - {2m \over r^2} + g^2 r^2 \prod_{i=1}^3H_i \,,
\qquad H_i=1+{{q_i}\over r^2}\,.
\label{5}
\ee
The three scalar fields $X_i$, subject to the constraint
$\prod_{i=1}^3 X_i=1$, have the potential
\be
{\cal V}=-\fft{g^2}{4\pi} \sum_{i} \fft1{X_i}\,.
\ee

   The relevant thermodynamic quantities are given by
\bea
E &=& \ft14\pi [ 3m + q_1+q_2+q_3]\,,\qquad 
  Q_i= \ft14 \pi \sqrt{q_i(q_i+2m)}\,,\qquad
 S= \ft12\pi^2\prod_i (r_+^2 +q_i)^{1/2}\,,\nn\\
T &=& \fft{f'(r_+)}{4\pi}\, \prod_i H_i^{-1/2}(r_+)\,,\qquad \Phi_i =
   \fft{\sqrt{q_i(q_i+2m)}}{(r_+^2 +q_i)}\,,
\eea
where the outer horizon is located at $r=r_+$, the largest root of $f(r_+)=0$.

   Substituting into the first law (\ref{first}) or the Smarr relation
(\ref{smarr}), we find that $\Theta$ is given by
\be
\Theta= -\fft{\pi r_+^4}{32}\, \prod_i H_i(r_+)\, \sum_j \fft1{H_j(r_+)}\,.
\ee
With the integral of the scalar potential defined by
\be
W= \int_{r_0}^{r_+} dr \int d\Omega_3 {\cal V}\sqrt{-g}\,,
\ee
where $r_0$ is taken to be the largest root of
\be
3 r_0^4 + 2 r_0^2 \sum_i q_i + \sum_{i<j} q_i q_j=0\,,
\ee
we find that $\Theta$ can again be written as
\be
\Theta = -\fft1{\Lambda}\, W\,.
\ee

\subsection{Charged AdS black holes in $D=7$}

   The metric, electromagnetic potentials, and scalars fields are
given by \cite{10author}
\bea
ds_7^2 &=& - (H_1 H_2)^{-4/3}\, f dt^2 + (H_1 H_2)^{1/5}\,
        \bigg(f^{-1} {dr^2} +
        r^2 d\Omega_{5}^2 \bigg) \,,\nn\\
A^i &=& \fft{\sqrt{q_i(q_i+ 2m)}}{(r^4 +q_i}\, dt\,,\qquad
X_i = H_i^{-1}\, (H_1 H_2)^{2/5}\,,
\eea
where
\be
f = 1 - {2m \over r^4} + g^2 r^2 H_1 H_2 \,,
\qquad H_i=1+{{q_i}\over r^4}\,.
\label{f7}
\ee
The two scalar fields $X_i$ have the potential
\be
{\cal V}=-\fft{g^2}{4\pi}\, (4 X_1 X_2 + 2 X_1^{-1} X_2^{-2} 
       + 2 X_2^{-1} X_1^{-2} -\ft12 (X_1 X_2)^{-4})\,.
\ee

The relevant thermodynamic quantities are
\bea
E &=& \ft18\pi^2 [ 5m + 2(q_1+q_2)]\,,\qquad
  Q_i= \ft14 \pi^2 \sqrt{q_i(q_i+2m)}\,,\qquad
 S= \ft14\pi^3 r_+\, \prod_i (r_+^4 +q_i)^{1/2}\,,\nn\\
T &=& \fft{f'(r_+)}{4\pi}\, (H_1(r_+) H_2(r_+))^{-1/2}\,,\qquad \Phi_i =
   \fft{\sqrt{q_i(q_i+2m)}}{(r_+^4 +q_i)}\,,
\eea
where the outer horizon is located at $r=r_+$, the largest root of $f(r_+)=0$.

   Substituting into the first law (\ref{first}) or the Smarr relation
(\ref{smarr}), we find that $\Theta$ is given by
\be
\Theta= -\fft{\pi^2 r_+^6}{96}\, [H_1(r_+) H_2(r_+) + 2H_1(r_+) + 2 H_2(r_+)]\,.
\ee
With the integral of the scalar potential defined by
\be
W= \int_{r_0}^{r_+} dr \int d\Omega_5 {\cal V}\sqrt{-g}\,,
\ee
where $r_0$ is taken to be the largest root of
\be
5 r_0^8 + 3(q_1+q_2) r_0^4  + q_1 q_2=0\,,
\ee
we find that $\Theta$ can again be written as
\be
\Theta = -\fft1{\Lambda}\, W\,.
\ee

\section{Rotating Black Holes}

\subsection{Kerr-AdS black holes in all dimensions}

   The Kerr-(A)dS solution in all dimensions, which generalises the 
asymptotically-flat rotating black hole solutions of \cite{myeper},
was obtained in \cite{gilupapo1,gilupapo2}. 
The metric obeys the vacuum Einstein equations $R_{\mu\nu}=-
(D-1)g^2 g_{\mu\nu}$.  In the `generalized' Boyer-Lindquist 
coordinates it takes the form
\begin{eqnarray}\n{MPC}
ds^2\!\!&=&\!\!-W(1+g^2r^2)d t^2+\frac{2m}{U} \Bigl(Wd t-
\sum_{i=1}^{N}\frac{a_i\mu_i^2d \phi_i}{\Xi_i}\Bigr)^{2} 
+\sum_{i=1}^{N}\frac{r^2+a_i^2}{\Xi_i}\,(\mu_i^2d \phi_i^2
+d \mu_i^2) \nonumber\\
\!\!&\ &\!\!+\frac{Ud r^2}{V-2m}-
\frac{g^2}{W(1+g^2r^2)}\,
\Bigl(\sum_{i=1}^N\frac{r^2+a_i^2}{\Xi_i}\,\mu_i d \mu_i+
\epsilon r^2\nu d \nu\Bigr)^2
+\epsilon r^2 d \nu^2\,,
\end{eqnarray} 
where
\begin{eqnarray}
W\!\!&\equiv&\!\!\sum_{i=1}^N\frac{\mu_i^2}{\Xi_i}+\epsilon \nu^2\,,\quad 
V\equiv r^{\epsilon-2}(1 +g^2 r^2)\prod_{i=1}^{N}(r^2+a_i^2)\,,\nonumber\\
U\!\!&\equiv &\!\!\frac{V}{1+g^2r^2}\,\Bigl(1-\sum_{i=1}^N\frac{a_i^2 \mu_i^2}{r^2+a_i^2}\Bigr)\,,\quad \Xi_i=1-g^2a_i^2\,.
\end{eqnarray}
Here $N\equiv [(D-1)/2]$, where $[A]$ means the integer part of $A$ and we 
have defined $\epsilon$ to be $1$ for $D$ even and $0$ for odd. The 
coordinates $\mu_i$ are not independent, but obey the constraint
\begin{equation}\label{constraint}
\sum_{i=1}^N\mu_{i}^2+\epsilon \nu^2=1\,.
\end{equation}
In the remainder of the paper, we shall not in general indicate the
range of the $i$ index in summations or products; it will always be understood
to be for $1\le i\le N$, with $N=(D-1)/2$ in odd dimensions, and 
$N=(D-2)/2$ in even dimensions.

The calculation of $\Theta$ is slightly different in the 
two cases that the dimension $D$ is odd or even.  We discuss
these cases in the following two subsections.

\subsubsection{Odd-dimensional Kerr-AdS black holes}

  Here, we take $D=2N+1$.  As discussed in \cite{GPP}, the various
thermodynamic quantities are given by
\bea\label{TDodd}
E&=& \fft{m \cA_{D-2}}{4\pi\prod_j\Xi_j}\, \Big(\sum_i\fft1{\Xi_i} -\ft12\Big)
\,,\quad J_i = \fft{m a_i \cA_{D-2}}{4\pi \Xi_i \prod_j\Xi_j}\,,\quad
S= \fft{\cA_{D-2}}{4r_+}\, \prod_i \fft{r_+^2+a_i^2}{\Xi_i}\,,\\
T&=& \fft{r_+(1+g^2 r_+^2)}{2\pi} \sum_i \fft1{r_+^2+a_i^2} -\fft1{2\pi r_+}
\,,\quad \Omega_i = \fft{(1+g^2 r_+^2) a_i}{r_+^2 + a_i^2}\,,
\eea
where $m$ and $a_i$ are the ``mass'' and the $N$ rotation parameters appearing
in the Kerr-AdS metrics, the summations and products are taken over 
$1\le i\le N$, the horizon radius is determined by the relation
\be
2m = \fft1{r_+^2}\, (1+g^2 r_+^2)\prod_i (r_+^2 + a_i^2)\,,
\ee
and the $\Xi_i$ are given by $\Xi_i=1-g^2 a_i^2$.  The quantity 
${\cal A}_{D-2}$ is the volume of the unit-radius $(D-2)$-sphere, and is
given by
\be
{\cal A}_{D-2} = \fft{2\pi^{(D-1)/2}}{\Gamma[(D-1)/2]}\,.
\ee
  
   After substituting into (\ref{first}) or (\ref{smarr}), 
we find that $\Theta$ is given by
\bea
\Theta\Lambda &=& \fft{m \cA_{D-2}}{8\pi\prod_j\Xi_j}
  \Big(\sum_i \fft1{\Xi_i} +\fft{D-3}{2} - \fft{D-2}{1+g^2 r_+^2}\Big)\,
\label{Thetaodd}\\
&=& \ft12 E -\fft{m(D-2){\cal A}_{D-2}}{16\pi\prod_i \Xi_i}\, 
  \fft{1-g^2 r_+^2}{1+g^2 r_+^2}\,.
\eea
This may in fact be written more simply if we introduce another quantity
$\Theta'$, such that
\be
\Theta = \Theta' - \fft1{2(D-1)}\, \sum_i a_i J_i\,,\label{redef2}
\ee
with $\Theta'$ being given by
\be
\Theta' = -\fft{(D-2) m \cA_{D-2}}{8\pi (D-1)\prod_i\Xi_i}\,
\fft{r_+^2}{1+g^2 r_+^2}= -\fft{(D-2)}{(D-1)} \, \fft{r_+ A}{16\pi}\,,
\label{thetapo}
\ee
where $A=4S$ is the area of the horizon.
Remarkably, $r_+ A$ is related to the spatial integral of $\sqrt{-g}$ up to
the horizon radius.  Specifically, we define
\be
V(r_+) = \int_{r_0}^{r_+} dr \int d\Omega \sqrt{-g}\,,\label{volint}
\ee
where $d\Omega$ denotes the integration over the coordinates parameterising
the $(D-2)$-sphere surfaces, and $r_0$ is given by $r_0^2=-a_{\rm min}^2$,
where $a_{\rm min}^2$ is the smallest amongst the values of the $a_i^2$.
(The $(D-2)$-spheres are not {\it round} spheres, of course.)
Using the expression for $\sqrt{-g}$ obtained in the appendix of
\cite{GPP}, we then find after some algebra that
\be
V(r_+) = \fft{r_+\, A}{D-1}\,.\label{volarea}
\ee
This therefore implies that
\be
\Theta' = -\fft{(D-2)}{16\pi}\, V(r_+)\,.
\ee
(Note that in performing the integration in eqn (\ref{volint}), it is
really more appropriate to use $x=r^2$ as the radial variable, since in
odd dimensions $r^2$ can be negative.)

   It is interesting also that $\Theta'$ can be obtained from a
Smarr relation if one works in a certain frame that is rotating at infinity.
Specifically, we have
\be
E' = \fft{D-2}{D-3}\, (TS + \Omega_i' J_i) + \Phi_\alpha Q_\alpha
   -\fft2{D-3}\, \Theta' \Lambda\,,\label{smarrp}
\ee
where $E'$ and $\Omega_i'$ are the energy and the angular velocities of
the horizon measured with respect to a frame defined 
by sending the azimuthal coordinates $\phi_i$ in the the
black-hole metrics to $\phi_i + a_i g^2 t$.  
This implies that\footnote{It should be noted, however,
that the thermodynamic variables $E'$ and $\Omega_i'$ do {\it not} satisfy
the first law of thermodynamics.  Thus, for example, if we hold $\Lambda$
fixed then $dE'$ is not equal to $T dS + \Omega_i' dJ_i$, and indeed,
the latter is not even an exact differential.  (See \cite{GPP} for  
a detailed discussion.)}
\bea
E'&=& E - g^2 \sum_i a_i J_i = \fft{(D-2) m \cA_{D-2}}{8\pi \prod_i\Xi_i}
 \,,\\
\Omega_i' &=& \Omega_i - a_i g^2 = \fft{a_i\, \Xi_i}{r_+^2+a_i^2}\,.
\eea
 
   The Einstein action in $D$ dimensions is (with $G=1$)
\be
{\cal I}_D = \fft1{16\pi}\, \int\sqrt{-g}[R- (D-2)\Lambda]\, d^Dx\,.
\ee
Thus if we define the potential $W$ to be
\be
W\equiv \fft{(D-2)\Lambda}{16\pi}\, 
    \int_{r_0}^{r_+} dr \int d\Omega \sqrt{-g}\,,
\ee
then we have
\be
\Theta' = -\fft{1}{\Lambda}\, W\,.\label{thetapot}
\ee

\subsubsection{Even-dimensional Kerr-AdS black holes}

  Here, we take $D=2N+2$.  As discussed in \cite{GPP}, the various
thermodynamic quantities are now given by
\bea\label{TDeven}
E&=& \fft{m \cA_{D-2}}{4\pi\prod_j\Xi_j}\,\sum_i\fft1{\Xi_i}
\,,\quad J_i = \fft{m a_i \cA_{D-2}}{4\pi \Xi_i \prod_j\Xi_j}\,,\quad
S= \ft14 \cA_{D-2}\, \prod_i \fft{r_+^2+a_i^2}{\Xi_i}\,,\\
T&=& \fft{r_+(1+g^2 r_+^2)}{2\pi} \sum_i \fft1{r_+^2+a_i^2} -
  \fft{1-g^2 r_+}{4\pi r_+}
\,,\quad \Omega_i = \fft{(1+g^2 r_+^2) a_i}{r_+^2 + a_i^2}\,,
\eea
and the location of the horizon is determined by the equation
\be
2m = \fft1{r_+}\, (1+g^2 r_+^2)\prod_i (r_+^2 + a_i^2)\,,
\ee
where the summations and products are over $1\le i \le N$.
We find from (\ref{first}) or from (\ref{smarr}) that 
\bea
\Theta\Lambda &=& \fft{m \cA_{D-2}}{8\pi\prod_j\Xi_j}
  \Big(\sum_i \fft1{\Xi_i} +\fft{D-2}{2} - \fft{D-2}{1+g^2 r_+^2}\Big)\,
\label{Thetaeven}\\
&=& \ft12 E - \fft{m(D-2){\cal A}_{D-2}}{16\pi\prod_i \Xi_i}\, 
 \fft{1-g^2 r_+^2}{1+g^2 r_+^2}\,.
\eea

   Again we find that $\Theta$ can be expressed more simply in the form
(\ref{redef2}), with $\Theta'$ given by (\ref{thetapo}). 
As in the odd-dimensional case, we again find that if we define a
volume ``inside the horizon'' as in (\ref{volint}), then the relation
(\ref{volarea}) again holds, and hence $\Theta'$ is again related to the
potential $W$ by equation (\ref{thetapot}).  The only difference from the
odd-dimensional case is that in the volume integral (\ref{volint}), the
lower limit for the radial integration should now be $r_0=0$.  (This
is really the same rule as is used in odd dimensions, since in 
even dimensions there is effectively a ``missing'' rotation parameter
that is equal to zero.)

\subsection{Rotating pairwise-equal 4-charge black hole in $D=4$ gauged
supergravity}\label{d4gaugesec}

  The metric for this black hole is obtained in \cite{chcvlupo4}. The
various thermodynamic quantities are given by \cite{timemachine}
\bea
E &=& \fft{m+q_1+q_2}{\Xi^2}\,,\qquad S= \fft{\pi (r_1 r_2+a^2)}{\Xi}\,,\qquad
J= \fft{a(m+q_1+q_2)}{\Xi^2}\,,\nn\\
Q_1&=&Q_2= \fft{\sqrt{q_1(q_1+m)}}{2\Xi}\,,\qquad
Q_3=Q_4=\fft{\sqrt{q_2(q_2+m)}}{2\Xi}\,,\nn\\
T&=& \fft{\Delta_r'}{4\pi(r_1 r_2+a^2)}\,,\qquad 
\Omega = \fft{a(1+g^2 r_1 r_2)}{r_1 r_2+a^2}\,,\nn\\
\Phi_1&=&\Phi_2= \fft{2r_1\sqrt{q_1(q_1+m)}}{r_1 r_2+a^2}\,,\quad
\Phi_3=\Phi_4= \fft{2r_2\sqrt{q_2(q_2+m)}}{r_1 r_2+a^2}\,,\label{2charge4}
\eea 
where $r_1=r+2q_1$, $r_2=r+2 q_2$,
\be
\Delta_r= r^2+a^2- 2m r + g^2 r_1 r_2(r_1 r_2+a^2)\,,
\ee
and all $r$-dependent quantities in (\ref{2charge4}) are evaluated at
the horizon radius $r_+$, determined as the largest root of 
$\Delta_r(r_+)=0$.

   Substituting into either (\ref{first}) or (\ref{smarr}), we can determine
$\Theta$.  As usual in rotating black holes, the expression is quite
complicated, and it is most elegantly expressed, {\it via} (\ref{redef2}),
in terms of $\Theta'$ defined in the rotating frame:
\be
\Theta = \Theta' - \ft16 a J\,,
\ee
where
\be
\Theta' = -\fft{r +q_1+q_2}{6\Xi} (r_1 r_2+a^2)\,,
\ee
(evaluated at $r=r_+$).

   There is a scalar potential in the four-dimensional gauged supergravity,
given by
\be
{\cal V} = -\fft{g^2}{16\pi}\, (4 + 2\cosh\varphi + e^\varphi\, \chi^2)\,,
\ee
and in the black hole solution we have \cite{chcvlupo4}
\be
e^\varphi = \fft{r_1^2 + a^2 \cos^2\theta}{r_1 r_2 + a^2\cos^2\theta}\,,
\qquad \chi= \fft{a(r_2-r_1)\cos\theta}{r_1^2 + a^2 \cos^2\theta}\,.
\ee
If we define
\be
U(r) = \int_0^{2\pi}d\phi \int_0^\pi d\theta\, {\cal V}\sqrt{-g}\,,
\ee
then we find that
\be
\fft{d\Theta'}{d r_+} = -\fft1{\Lambda}\, U(r_+)\,.
\ee
In integral form, if we define the potential term
\be
W = \int_{r_0}^{r_+} dr \int_0^{2\pi} d\phi \int_0^\pi d\theta 
 {\cal V} \sqrt{-g}\,,
\ee
then 
\be
\Theta' = -\fft1{\Lambda}\, W\,,
\ee
where the lower limit of integration is taken to be 
\be
r_0=-q_1-q_2 +
    \sqrt{(q_1-q_2)^2 -a^2}\,.
\ee

\subsection{Charged rotating black hole in minimal $D=5$ gauged supergravity}
\label{d5gaugesec}

  The metric for this black hole is obtained in \cite{chcvlupo}.  It has
the thermodynamic quantities
\bea
E&=& \fft{m\pi(2\Xi_a + 2\Xi_b - \Xi_a \Xi_b)+ 2\pi q a b g^2(\Xi_a+\Xi_b)}{
 4 \Xi_a^2 \Xi_b^2}\,,\qquad S=\fft{\pi^2[(r_+^2+a^2)(r_+^2+b^2)+a b q]}{
         2\Xi_a \Xi_b\, r_+}\,,\nn\\
J_a &=& \fft{\pi(2am+qb(1+a^2g^2)]}{4\Xi_a^2 \Xi_b}\,,\qquad
J_b = \fft{\pi(2bm+qa(1+b^2g^2)]}{4\Xi_b^2 \Xi_a}\,,\qquad
Q= \fft{\sqrt3\, \pi q}{4\Xi_a\Xi_b}\,,\nn\\
T&=&\fft{r_+^4[1+g^2(r_+^2+a^2+b^2)] -(ab+q)^2}{
  2\pi r_+[(r_+^2+a^2)(r_+^2+b^2)+abq]}\,,\qquad
\Phi= \fft{\sqrt3 \, q r_+^2}{(r_+^2+a^2)(r_+^2+b^2)+abq}\,,\nn\\
\Omega_a &=& \fft{a(r_+^2+b^2)(1+g^2 r_+^2)+b q}{
   (r_+^2+a^2)(r_+^2+b^2)+abq}\,,\qquad
\Omega_b = \fft{b(r_+^2+a^2)(1+g^2 r_+^2)+a q}{
   (r_+^2+a^2)(r_+^2+b^2)+abq}\,,
\eea
where the location of the horizon is determined by the equation
\be
2m = \fft{(r_+^2+a^2)(r_+^2+b^2)(1+g^2 r_+^2)+q^2+ 2abq}{r_+^2}\,.
\ee

  From (\ref{first}) or from (\ref{smarr}) we find that 
\bea
\Theta &=&\Theta_0 -\fft{abq\pi}{16\Xi_a^2\Xi_b^2 r_+^2}\,
 \Big[ 2r_+^2 + a^2+b^2-g^2(r_+^2(a^2+b^2)+ 2 a^2 b^2)\Big]\nn\\
&& - \fft{\pi q^2(a^2+b^2-2 a^2 b^2 g^2)}{32\Xi_a^2\Xi_b^2 r_+^2}\,,
\eea
where $\Theta_0$ is the value for five-dimensional Kerr-AdS, as
given in (\ref{Thetaodd}) for $D=5$.  As in the Kerr-AdS examples, 
the quantity $\Theta'$ evaluated in the asymptotically rotating frame,
and defined by (\ref{redef2}), is much simpler, and is given in this case by
\be
\Theta' = -\fft{\pi}{32\Xi_a\Xi_b}\, [3(r_+^2+a^2)(r_+^2+b^2)+2abq]\,.
\ee

 The metric in \cite{chcvlupo} has 
\be
\sqrt{-g} = \fft{r\sin\theta\cos\theta\, (r^2+ a^2 \cos^2\theta +
  b^2\sin^2\theta)}{\Xi_a\Xi_b}\,,
\ee
and hence if we define
\be
U(r)\equiv -\fft{3g^2}{4\pi}\, \int_0^{2\pi} d\phi \int_0^{2\pi} d\psi 
\int_0^{\ft12\pi} d\theta
\sqrt{-g} = \fft{3 g^2\pi r(2r^2+a^2+b^2)}{4\Xi_a\Xi_b}\,,
\ee
(where $-3g^2/(4\pi)$ is the coefficient of the cosmological term in 
the Lagrangian),
then we see that
\be
\fft{d\Theta'}{dr_+}= -\fft1{\Lambda}\, U(r_+)\,.
\ee
To integrate this we introduce the radial variable $x=r^2$, and integrate from
$x=x_0$ to $x=r_+^2$, where $x_0$ is the less negative of the two 
possibilities
\be
x_0 = -\ft12(a^2+b^2) \pm \ft12\sqrt{(a^2-b^2)^2 - \ft83 a b q}\,.
\ee

\section{Reverse  Isoperimetric Inequality}

   The isoperimetric inequality for
the volume $V$ of a connected domain   
in Euclidean space ${\Bbb E} ^{D-1}$ whose area
is $A$ states that
\ben
\Bigl( \frac{(D-1)V}{{\cal A}_{D-2} } \Bigr ) ^{D-2 } \le
\Bigl( \frac{A} {{\cal A}_{D-2} } \Bigr ) ^ {D-1 }  
\label{iso} 
\een  
with equality if and only if the domain is a standard round ball.  
Thus we may restate the inequality as $R\le 1$, where we define
\be
R\equiv \Bigl( \frac{(D-1)V}{{\cal A}_{D-2} } \Bigr )^{\fft1{D-1}}\,
  \Bigl(\fft{{\cal A}_{D-2}}{A}\Bigr)^{\fft1{D-2}}\,.\label{Rdef}
\ee

   It is interesting to examine whether or not the area of the black
hole horizon and the ``volume'' defined via either $\Theta$ or $\Theta'$ satisfy
the isoperimetric inequality.  Let us first consider the case of
electrically neutral black holes; i.e., the rotating Kerr-AdS black holes
in arbitrary dimensions.  Intriguingly, we find that if we use the
quantity $\Theta'$ to define the volume of the black hole, then the
isoperimetric inequality is always satisfied in Kerr-AdS, with equality
being attained for the non-rotating Schwarzschild-AdS limit.  If, on the
other hand, we use the quantity $\Theta$, which arises naturally from
thermodynamic considerations, to define the volume, then the opposite
is true, and the isoperimetric inequality is always violated, except in the
non-rotating limit.

\subsection{Isoperimetric inequality for the $\Theta'$ volume}

   For the Kerr-AdS metrics, if $A$ is the area of the event horizon, 
then in all cases
\ben
V' =    - \frac{16 \pi }{D-2} \Theta ^\prime =  
\frac{r_+ A}{D-1} \,, \label{Vprimekads}
\een
and if $D$ is odd
\ben
A= \frac{ {\cal A}_{D-2}}{r_+} \prod_i \frac{r_+^2 + a_i^2 } {\Xi_i}\,,  
\label{Aodd}
\een
whilst if $D$ is even
\ben
A= {\cal A}_{D-2} \prod_i \frac{r_+^2 + a_i^2 } {\Xi_i} \,.
\label{Aeven}
\een

    A simple calculation shows that in both odd and even dimensions, 
$R'$ defined by (\ref{Rdef}), and using the volume $V'$,  given by
\be
R' = \prod_i \Big(\fft{1+a_i^2/r_+^2}{\Xi_i}\Big)^{-\fft1{(D-1)(D-2)}}\,.
\ee
Since $\Xi_i=1-g^2 a_i^2 \le 1$ for each $i$, it is evident that $R'\le1$,
with equality when all $a_i$ vanish.

Thus remarkably, 
the geometrical $V'$
and the surface area $A$ of the black hole
satisfy  the standard isoperimetric inequality for a ball in flat
Euclidean space  ${\Bbb E} ^{D-1}$. There is an obvious analogy here 
with the liquid drop model, which regards a nucleus as a ball
of incompressible fluid, whose volume is thus fixed.
If the energy is solely due to positive surface  tension,  then
the configuration which minimizes the energy is spherical.
  
\subsection{Reverse isoperimetric inequality for the $\Theta$ volume}

    We saw in equation (\ref{redef2}) that the thermodynamic quantity 
$\Theta$ in Kerr-AdS is more negative than $\Theta'$, and hence it
follows that the associated volume $V$ is larger than $V'$.  In fact, from
(\ref{Vprimekads}) and (\ref{redef2}) we find that
\be
V= \fft{r_+\, A}{(D-1)}\, \Big[1 + \fft{(1+g^2 r_+^2)}{(D-2) r_+^2}\,
      \sum_i\fft{a_i^2}{\Xi_i}\Big]\,.\label{Vkads}
\ee
This suggests
the possibility that although $V'$ and $A$ satisfy the isoperimetric
inequality, as we saw above, it might be that the volume $V$ and the area
$A$ could violate it in Kerr-AdS black holes.  This is indeed exactly
what we find.  Since, as it turns out, this violation 
 seems to be a universal property,
for all rotating and/or charged black holes, we may elevate this to the
status of a conjecture in its own right.  Thus we make the conjecture
that the ratio $R$ defined in (\ref{Rdef}) actually satisfies
the {\it Reverse Isoperimetric Inequality}
\be
  R \ge 1\label{isoa}
\ee
for all black holes, if one uses the ``thermodynamic'' definition of
the volume $V$.   We now demonstrate the validity of the conjecture 
for a variety of black hole solutions.

\subsubsection{Kerr-AdS}

   Defining the (necessarily non-negative) dimensionless quantity
\be
z= \fft{(1+g^2 r_+^2)}{r_+^2}\,\sum_i\fft{a_i^2}{\Xi_i}\,,\label{zdef}
\ee
we consider $R^{D-1}$, where $R$ is given by (\ref{Rdef}), and
observe that in odd dimensions
\bea
R^{D-1} &=&r_+\, \Big[1+ \fft{z}{D-2}\Big]\, \Big[\fft1{r_+}\, \prod_i 
   \fft{(r_+^2+a_i^2)}{\Xi_i}\Big]^{-\fft1{D-2}}\,\nn\\
&=&  \Big[1+ \fft{z}{D-2}\Big]\, \Big[\prod_i
   \fft{(r_+^2+a_i^2)}{r_+^2\, \Xi_i}\Big]^{-\fft1{D-2}}\,\nn\\
&\ge& \Big[1+ \fft{z}{D-2}\Big]\, \Big[\fft{2}{D-1}\Big(
\sum_i\fft1{\Xi_i} + \sum_i \fft{a_i^2}{r_+^2\, \Xi_i}\Big)
       \Big]^{-\fft{(D-1)}{2(D-2)}} \,\nn\\
&=& \Big[1+ \fft{z}{D-2}\Big]\, \Big[1+ \fft{2z}{D-1}
   \Big]^{-\fft{(D-1)}{2(D-2)}}\equiv F(z)\,,
\eea
where the inequality follows from $(\prod_i x_i)^{1/N} \le (1/N)\sum_i x_i$
for non-negative quantities $x_i$.  

   Noting that $F(0)=1$, and that
\be
\fft{d\log F(z)}{dz}= \fft{(D-3)\, z}{(D-2)(D-2+z)(D-1+2z)}\,,
\ee
which is positive for non-negative $z$ in $D>3$ dimensions, it follows
that $F(z)\ge1$, and hence the reverse isoperimetric inequality (\ref{isoa})
is satisfied by all odd-dimensional Kerr-AdS black holes.

  In even dimensions the calculation is rather similar, since now we have
\bea
R^{D-1} &=&r_+\, \Big[1+ \fft{z}{D-2}\Big]\, \Big[\prod_i
   \fft{(r_+^2+a_i^2)}{\Xi_i}\Big]^{-\fft1{D-2}}\,\nn\\
&=&  \Big[1+ \fft{z}{D-2}\Big]\, \Big[\prod_i
   \fft{(r_+^2+a_i^2)}{r_+^2\, \Xi_i}\Big]^{-\fft1{D-2}}\,\nn\\
&\ge& \Big[1+ \fft{z}{D-2}\Big]\, \Big[\fft{2}{D-2}\Big(
\sum_i\fft1{\Xi_i} + \sum_i \fft{a_i^2}{r_+^2\, \Xi_i}\Big)
       \Big]^{-\ft12} \,\nn\\
&=& \Big[1+ \fft{z}{D-2}\Big]\, \Big[1+ \fft{2z}{D-2}
   \Big]^{-\ft12}\equiv G(z)\,.
\eea
Thus $G(0)=1$ and $d\log G(z)/dz\ge0$, and so again we conclude that
$R\ge1$.  Thus the reverse isoperimetric inequality holds for even-dimensional
Kerr-AdS black holes also.

\subsubsection{Charged static black holes}

  All of the charged static black hole solutions in gauged supergravity
satisfy the reverse isoperimetric inequality also.  There is no distinction
between the $V$ and $V'$ volumes in this case, since there is no rotation.
Consider, for example, the
4-charge solution given in section \ref{4chargesec}. The volume and area
are given by
\be
V= \ft13 \pi\, \sum_i\fft1{r_++q_i}\, \prod_j (r_++q_j)\,,\qquad
A= 4\pi \prod_i (r_++q_i)\,,
\ee
and so from (\ref{Rdef}) we have
\be
R^3 = \ft14 \sum_i\fft1{r_++q_i}\, \prod_j(r_+ +q_j)\,,
\ee
and so using the inequality
\be
\prod_i (r_+ + q_i)^{-\ft14}  \le \ft14\sum_i\fft1{r_+ + q_i}\,,
\ee
we see that $R\ge 1$.

  Very similar calculations show that the inequality $R\ge 1$ holds for
the static charged black holes in $D=5$ and $D=7$ also.

\subsubsection{Charged rotating black holes}

   We have verified explicitly that $R\ge 1$ for the
rotating black hole in four-dimensional gauged supergravity with pairwise
equal charges (described in section \ref{d4gaugesec}), and also for the 
charged rotating black hole in five-dimensional ungauged minimal supergravity
(i.e. setting $g=0$ in the solution described in section \ref{d5gaugesec}).
In each case, the calculations are quite complicated, and we shall not
present them here.

   In the case of the rotating black hole in five-dimensional {\it gauged}
minimal supergravity, we have constructed an analytical proof that
$R\ge1$ in the case that the product $a b q$ is non-negative.   
Numerical investigations indicate that $R\ge 1$ also if $abq$ is negative.

  It is worth remarking that whilst we can obtain an expression for the 
volume $V$ of an asymptotically flat  black hole in ungauged supergravity
(or with zero cosmological constant) by sending $g\rightarrow 0$
or $\Lambda\rightarrow 0$ in the expressions obtained for an asymptotically
AdS black hole, we do not have an intrinsic way in general of defining
$V$ for an asymptotically flat black hole if the more general 
asymptotically AdS solution is not itself known.

The dependence of volume on $g$ is smooth; there are no discontinuities 
for $g\to 0$ or in the large to small black hole transition. To illustrate this point 
we display the $V=V(g)$ dependence for a Kerr-AdS black hole of 
fixed mass in Fig. 1.  

\begin{figure}
\begin{center}
\rotatebox{-90}{
\includegraphics[width=0.55\textwidth,height=0.50\textheight]{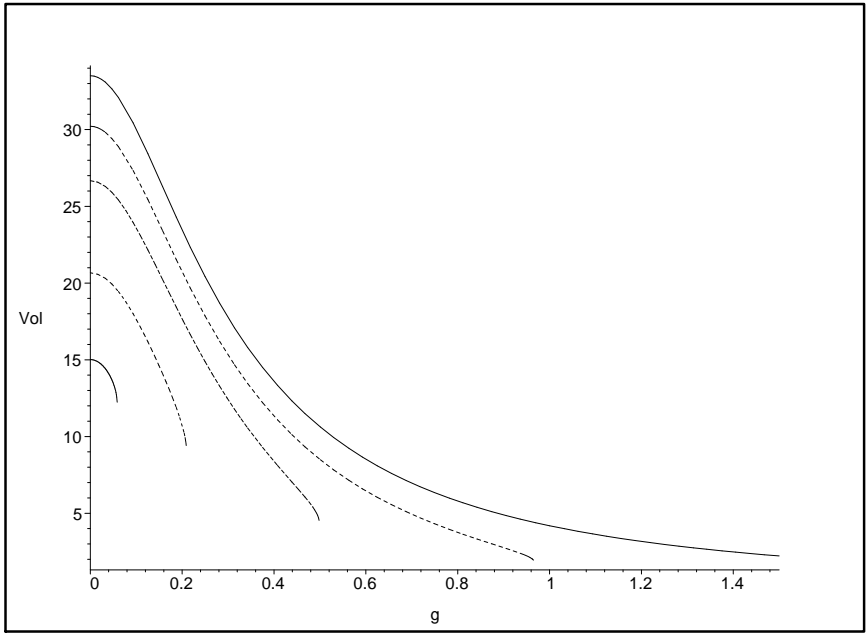}
}
\caption{{\bf Thermodynamic volume of the Kerr-AdS black hole.} The graph 
displays the dependence of $V$ on gauge coupling $g$, $\Lambda=-3g^2$, for 
various rotation parameteres, while we keep the total gravitational enthalpy 
fixed, $E=1$
 $(J=a)$. The upper curve represents Schwarzschild-AdS $(a=0$), the 
lower curves, in descending order, correspond to Kerr-AdS with
$a=0.5$, $a=0.7$, $a=0.9$ 
and $a=0.99$, respectively. Obviously, the smooth limit exists 
for $g\to 0$, the volume is smooth also in the transition between large 
and small black holes.}  
\end{center}
\end{figure}

  We have not checked our Reverse Isoperimetric Conjecture for all the
known examples of charged rotating black holes in gauged supergravities.
We have, however, examined the recent construction in \cite{4chow} of
the rotating black hole in four-dimensional maximal gauged supergravity
with two zero charges and the other two freely specifiable.  With non-zero
gauge coupling the complexity of the metric has so far prevented us from
obtaining an analytic proof, but the indications from numerical analysis are
that the conjecture is satisfied.  The expression for the thermodynamic
volume $V$ is much simpler in the limit that $g=0$, and in this case we
have been able to show analytically that the reverse isoperimetric conjecture
is satisfied.

  We have evaluated the volume for all solutions known to us in all
dimensions $D\leq 7$.  When $g\ne0$, the
volume may be obtained from the Smarr formula (\ref{smarr}) by dividing
by $\Lambda \sim -g^2$.  If  $D\le 6$, the numerator is always found to be
proportional to $g^2$, and hence a smooth limit exists as $g$ tends to zero.
In $D=7$, however, the numerator contains in addition 
terms proportional to $g$ (times the product of the three rotation
parameters $a_i$), and hence the $g\rightarrow 0$ limit diverges if all the
$a_i$ are nonvanishing.
The fact that the $D=7$ solutions \cite{chow7} 
are not invariant under $g\rightarrow -g$
may be traced back to the self-duality constraint for the 3-form gauge
potential in the seven-dimensional gauged supergravity theory (see for example, 
equation (3.8) in \cite{chow7}), since this equation contains a term
linear in $g$.

\section{Komar Integration, Smarr Formula and Killing Potentials for 
Kerr-AdS Black Holes}

\subsection{Komar derivation of the Smarr relation}

   In a $D$-dimensional 
stationary, axisymmetric black hole spacetime, let $\Sigma$ denote 
a spacelike hypersurface that intersects ${\cal H}^+$, a Killing horizon 
of $\xi=k+\Omega_i m_i$,  in a $(D-2)$-sphere $H$. Here, $k$ is a
Killing vector that is timelike at infinity, $m_i$ are $U(1)$ 
Killing fields that generate rotations in the orthogonal spatial 2-planes,
and $\Omega_i$ are the corresponding angular velocities of the horizon.
Since any Killing vector satisfies $\nabla_\mu K^\mu=0$ and $\square K_\mu
+ R_{\mu\nu} K^\nu=0$ it follows that if the metric is Ricci flat, 
corresponding to the case of an asymptotically-flat black hole, then
$d{*d}\xi=0$, and hence
\be
0=\int_{\Sigma} d{*d}\xi =\int_{\del\Sigma} {*d}\xi =
  \int_{S_\infty} {*d}\xi -\int_H {*d}\xi  \,,\label{afk}
\ee
where $S_\infty$ denotes the sphere at infinity.  One can show that
the Komar integrals constructed using the Killing vectors $k$ and $m_i$ 
give the energy and angular momenta of the asymptotically-flat black hole 
\be
E= -\fft{(D-2)}{16\pi (D-3)}\, \int_{S_\infty} {*d}k\,,\qquad
J_i = \fft1{16\pi}\, \int_{S_\infty} {*d}m_i\,,\label{komarflat}
\ee
while the integral of ${*d}\xi$ over the horizon gives
\be
\fft1{16\pi} \,\int_H {*d}\xi= \fft{\kappa\, A}{8\pi}= T S\,,
\ee
and so from (\ref{afk}) one obtains the Smarr relation
\be
E= \fft{(D-2)}{(D-3)}\, \Big( \fft{\kappa \, A}{8\pi} + \Omega_i J_i\Big) =
 \fft{(D-2)}{(D-3)}\, (TS +\Omega_i J_i)
\ee
for an asymptotically-flat black hole.

   If the cosmological constant is negative rather than zero, then the above
Komar derivation of the Smarr relation requires modification.  Following
the arguments in \cite{bazzyl,Kastor,KRT}, one may note that since 
any Killing vector
satisfies $d{*K}=0$, there must always exist, locally, a 2-form
{\it Killing potential}
$\omega_K$ such that $K$ may be written as $K= {*d}{*\omega_K}$.  
In view of
the fact that with $R_{\mu\nu}=\Lambda g_{\mu\nu}$ we now have
$d{*d}\xi +2\Lambda\, {*\xi}=0$, and it follows that 
\be
d{*d}\xi + 2\Lambda\, d{*\omega_\xi}=0\,.
\ee
By integrating this over $\Sigma$, one thereby obtains
\bea
0&=&\int_\Sigma (d{*d}\xi + 2\Lambda\, d{*\omega_\xi})=
\int_{\del\Sigma} ({*d}\xi + 2\Lambda {*\omega_\xi})\nn\\
&=& \int_{S_\infty} ({*dk} + 2\Lambda {*\omega_\xi}) +
  \Omega_i \int_{S_\infty} {*d}m_i   -
    \int_H {*d}\xi - 2\Lambda\, \int_H {*\omega_\xi}\,.\label{komarads}
\eea

Of course the Killing potential $\omega_\xi$ is not unique; one may add
any co-closed 2-form $\nu$ to $\omega_\xi$.  If $\nu$ is co-exact,
$\nu={*d}{*\eta}$ for any 3-form $\eta$, then the integrals of ${*\omega_\xi}$
in (\ref{komarads}) will be unaltered, since
\be
\int_{S_\infty} d{*\eta} = \int_{\del S_\infty} {*\eta}=0\,,\qquad
\int_{H} d{*\eta} = \int_{\del H} {*\eta}=0\,.
\ee
However, if $\nu$ is co-closed
but not co-exact, each of the integrals $\int_{S_\infty} {*\omega_\xi}$
and $\int_H {*\omega_\xi}$ will be separately changed by the addition of $\nu$,
although their difference will be unaltered, since
\be
\int_{S_\infty} {*\nu} - \int_H {*\nu} = \int_\Sigma d{*\nu}=0\,.\label{nutot}
\ee

   By analogy with the asymptotically-flat case we discussed above, one
would like to interpret the integrals over $S_\infty$ in 
(\ref{komarads}) as being proportional respectively 
to the energy and the angular momenta of the black hole.  Indeed,
one again finds that the integrals of ${*d}m_i$ give the angular momenta,
as in (\ref{komarflat}).  The integral $\int_{S_\infty} {*d}k$ by itself
now diverges, as does $\int_{S_\infty} {*\omega_\xi}$, but remarkably, 
the combination $\int_{S_\infty} ({*dk} + 2\Lambda {*\omega_\xi})$ turns
out to be finite.  Since, however, as we remarked above, its value is 
altered if one exploits the gauge freedom to add a co-closed 2-form $\nu$
to $\omega_\xi$, one cannot use 
$\int_{S_\infty} ({*dk} + 2\Lambda {*\omega_\xi})$ to provide an
unambiguous definition of the energy of the black hole.  The best that can be
done is to make a gauge choice for $\omega_\xi$ such that 
\be
E= -\fft{(D-2)}{16\pi (D-3)}\, 
   \int_{S_\infty} ({*dk} + 2\Lambda {*\omega_\xi})\label{komaradsmass}
\ee
yields the true mass $E$ of the black hole, which itself is determined
by other means.  

    The easiest and most reliable way of calculating the 
mass of an asymptotically AdS black hole is by means of the conformal
definition of Ashtekar, Magnon and Das (AMD) \cite{ashmag,ashdas}.  This has the
great advantage over other methods, such as that of Abbott and Deser
\cite{abodes}, that it involves an integration at infinity of a finite quantity,
computed from the Weyl tensor, that does not require any infinite
subtraction of a pure AdS background.   The AMD mass for the Kerr-AdS black
hole in arbitrary dimension was calculated in \cite{GPP}, and it
was shown to be consistent with the first law of thermodynamics.

   Having chosen a gauge for $\omega_\xi$ for which the integration in
(\ref{komaradsmass}) yields the AMD mass $E$, the remaining integrals in
(\ref{komarads}) can be evaluated.  Defining
\be
\Theta = \fft{(D-2)}{16\pi}\, \int_H {*\omega_\xi}\,,\label{Thetaint}
\ee
we recover precisely the Smarr relation (\ref{smarr}) for the uncharged case,
\be
E = \fft{D-2}{D-3}\, (TS + \Omega_i J_i)
   -\fft2{D-3}\, \Theta \Lambda\,.\label{smarr2}
\ee

\subsection{Killing potentials from the conformal Killing-Yano tensor}
\label{PCKYsec}

   In this subsection we review the work of \cite{KrtousEtal:2007jhep},
which shows how one may construct the towers of hidden and explicit symmetries
of a spacetime that admits a {\it Principal Conformal Killing-Yano 
(PCKY) tensor}.  In this discussion we closely follow 
\cite{FrolovKubiznak:2008}, and then we present  
a new method for constructing the Killing potentials for the Killing vectors.

    The PCKY tensor ${h}$ is a non-degenerate closed conformal 
Killing-Yano 2-form \cite{KrtousEtal:2007jhep}. 
This means that there exists a 1-form $\eta$ such that
\be\label{PCKY}
\nabla_\mu h_{\nu\rho} =2 g_{\mu[\nu}\, \eta_{\rho]}\,.
\ee 
The condition of non-degeneracy means that at a generic point of the 
manifold, the skew-symmetric matrix $h_{\mu\nu}$ 
has the maximum possible (matrix) rank, and that the eigenvalues of 
$h_{\mu\nu}$ 
are functionally independent in some spacetime domain. 
The equation \eqref{PCKY} implies
\be\label{xi}
{dh}=0\,,\qquad {\eta}=\frac{1}{D-1}\,{*d*h}\,.
\ee
This means that there exists a 1-form PCKY potential $b$, such that
\be
{h}={db}\,.
\ee
The 1-form ${\eta}$ associated with ${h}$ is called {\em primary}, and 
turns out to be a Killing 1-form.

   The PCKY tensor generates a tower of closed conformal Killing-Yano 
(CKY) tensors \cite{KrtousEtal:2007jhep}
\be\label{hj}
{h}^{(j)}\equiv {h}^{\wedge j}=\underbrace{{h}\wedge \ldots \wedge
{h}}_{\mbox{\tiny{total of $j$ factors}}}\, .
\ee 
The CKY tensor ${h}^{(j)}$ is a $(2j)$-form, and in particular 
${h}^{(1)}={h}$.
Since ${h}$ is non-degenerate, one has a
set of $N+\eps$ nonvanishing closed CKY tensors in dimension $D=2N+1+\eps$, 
where $\eps=0$ in odd dimensions and $\eps=1$ in even dimensions. 
In an even-dimensional
spacetime,  ${h}^{(N+1)}$ is proportional to the totally antisymmetric
tensor, whereas it is dual to a Killing vector in odd dimensions. In both 
cases 
such a CKY tensor is trivial, and can be excluded from the tower of 
hidden symmetries. 
Therefore we take $j=1,\dots, N-1+\eps$.

    The CKY tensors \eqref{hj} can be generated from the potentials ${b}^{(j)}$,
\be\label{bj}
{b}^{(j)}={b}\wedge {h}^{\wedge (j-1)}\,,\quad
{h}^{(j)}={d}{b}^{(j)}\,.
\ee 
For each $(2j)$-form ${h}^{(j)}$, its Hodge dual is a
$(D-2j)$-form, denoted by
\be\label{fj}
{f}^{(j)}={*}{h}^{(j)}\, .
\ee
In their turn, these tensors give rise to the Killing tensors
${K}^{(j)}$,
\be\n{Kj}
K^{(j)}_{\mu\nu}\equiv{1\over (D-2j-1)!(j!)^2} 
f^{(j)}{}_{\mu \rho_1\ldots \rho_{D-2j-1}}\,
f^{(j)}{}_\nu{}^{\rho_1\ldots \rho_{D-2j-1}}\, .
\ee
(The coefficient in this definition \eq{Kj} is a convenient choice in 
the canonical basis, see \cite{FrolovKubiznak:2008}.)
The metric itself trivially satisfies the conditions for a Killing tensor,
and it is convenient to define 
$K^{(0)}_{\mu\nu}=-g_{\mu\nu}$, extending the range of the $j$ index
so that $j=0,\dots , N-1+\eps$.

     The PCKY tensor also naturally generates $(N+1)$ vectors 
${\eta}^{(k)}$ $(k=0,\dots,N)$ which turn out to be the independent 
commuting Killing vector fields. These are given as 
\begin{equation}\label{etaj}
\eta^{(j)\,\mu}= K^{(j)}\,\!^\mu_{\ \nu}\, \eta^\nu\,,
\qquad j=0,\dots ,N-1+\eps\,,
\end{equation}
where $\eta^\mu$ is the Killing vector given by (\ref{xi}).
In odd dimensions the last Killing vector is in the tower is given by
the $N$-th Killing-Yano tensor 
\be\label{xin}
{\eta}^{(N)}=-\fft1{N!}\, {f}^{(N)}\,.
\ee

The {\em canonical spacetimes} with all these symmetries were 
constructed in \cite{HouriEtal:2009a,KrtousEtal:2008}.  When the Einstein 
equation is imposed, they are the general Kerr-NUT-AdS spacetimes 
constructed in \cite{chlupo}.

\subsubsection{Killing potentials}

   We shall now show how the PCKY tensor may be used in order to
 construct the Killing potentials for the Killing vectors. We define the 
following 2-forms, for $j=0,\dots N-1+\eps$:
\be\label{omegai}
\omega^{(j)}_{\mu\nu}=\frac{1}{D-2j-1}K^{(j)}_{\mu\rho}h^\rho{}_\nu\,,\qquad 
\omega^{(N)}=\frac{\sqrt{-c}}{N!}{*b}^{(N)}\,,
\ee
where the second expression, for $\omega^{(N)}$, applies only in odd 
dimensions. $\sqrt{-c}$ is some appropriately chosen constant
(see \cite{FrolovKubiznak:2008}). 
It is easy to verify (for example in the canonical basis) that these are 
Killing potentials for the previously-constructed Killing fields, i.e, we have
\be
\eta^{(i)}={*d}{*\omega}^{(i)}\,.
\ee
Note that although in odd dimensions the gauge freedom $b\to b+d\lambda$ 
affects $\omega^{(N)}$,
\be\label{freedom}
\omega^{(N)}\to \omega^{(N)}+\frac{\sqrt{-c}}{N!}\, {*d}(\lambda h^{(N-1)})\,,
\ee
its divergence
${*d}{* \omega^{(N)}}$ remains unchanged. Since any Killing vector 
$\xi$ in a canonical spacetime is a linear combination of 
the $\eta^{(i)}$, of the form $\xi=\sum_{i=0}^{N} c_i\eta^{(i)}$, the 
problem of finding its Killing potential reduces to the algebraic problem of 
finding the constant coefficients $c_i(\xi)$ of this expansion:
\be\label{co}
\xi={*d}{*\omega_\xi}\,,\qquad \omega_\xi=\sum_{i=0}^{N}c_i(\xi)\, 
    \omega^{(i)}\,,
\ee
where $\omega^{(i)}$ are given by \eq{omegai}.

\subsection{Kerr-AdS black holes}
\label{KPsec}

    The Kerr-AdS black hole metrics \eq{MPC} possess a closed 
conformal Killing-Yano 2-form $h$ \cite{KubiznakFrolov:2007} which can be 
derived from the potential $b$, $h=db$, given by 
\ba
{b}\!\!&=&\!\!\frac{1}{2}\biggl\{\Bigl[r^2+\sum_{i=1}^N 
a_i^2\mu_i^2\bigl(1 + g^2\, \frac{r^2+a_i^2}{\Xi_i}\bigr)\Bigr] d t-
\sum_{i=1}^N a_i\mu_i^2\frac{r^2+a_i^2}{\Xi_i} d \phi_i\biggr\}\,.
\label{b_MPlambda}
\ea
The 2-form $h$ is non-degenerate, i.e., it is a PCKY tensor when all 
rotations $a_i$ are non-zero and distinct. In that case any Killing vector 
of the spacetime is a linear combination of the (independent)
Killing fields $\eta^{(i)}$, and its Killing potential is given by \eq{co}, 
where in odd dimensions we identify the constant 
$\sqrt{-c}=\prod_{i=1}^N a_i$.\footnote{If the $a_i$ are not distinct or if
some of them vanish, then $h$ is degenerate. In such a case one does not 
recover all the Killing fields of the spacetime by the 
construction \eq{etaj} and \eq{xin}. However, the formula for the Killing 
potential $\omega_\xi$ obtained in the next section, eqn \eq{omegaxi}, 
still applies.}

   The outer Killing horizon of the Kerr-AdS metric (\ref{MPC}) is 
located at $r=r_+$, the largest root of $V(r_+)-2m=0$.  It is a Killing 
horizon for the  Killing field 
\be
\xi=\partial_t+\Omega_i \partial_{\phi_i}\,,\qquad 
\Omega_i=\frac{a_i(1+g^2r_+^2)}{r_+^2+a_i^2}\,.
\ee
The Killing potential $\omega_\xi$, \eq{co}, 
now reads
\be\label{omegaxi}
\omega_\xi=\frac{r_+^{2N}}{\prod_{i=1}^N(r_+^2+a_i^2)}\,
\sum_{j=0}^{N}\frac{1}{r_+^{2j}}\,\omega^{(j)}\,.
\ee

  Before using $\omega_\xi$ in (\ref{komarads}) to derive
the Smarr relation, we must first consider the gauge freedom to add to it
a non-trivial co-closed 2-form $\nu$.  Since co-closure, or divergence 
freedom, can be written as
\be
\del_\mu(\sqrt{-g}\, \nu^{\mu\nu})=0\,,
\ee
it is clear that a co-closed $\nu$ is obtained if we take all its contravariant
components to vanish except for $\nu^{tr} =\hbox{constant}/\sqrt{-g}$.  
This is equivalent to the statement that
\be
{*\nu}= \alpha\, \Omega_{D-2}\,,
\ee
where $\alpha$ is a constant and $\Omega_{D-2}$ is the volume element of
the unit $(D-2)$-sphere.  Evaluating (\ref{komaradsmass}) with
$\omega_\xi$ given by (\ref{omegaxi}) plus $\nu$,
\be
\omega_\xi\longrightarrow \tilde\omega_\xi=
\omega_\xi - \alpha\, {*\Omega_{D-2}}\,,
\ee
we find that in order for (\ref{komaradsmass}) to produce the correct
AMD mass for the Kerr-AdS black holes, we must choose
\be
\alpha = -\fft{2m}{(D-1)(D-2)\, (\prod_j \Xi_j)}\, \sum_i \fft{a_i^2}{\Xi_i}\,.
\ee
Using $\tilde\omega_\xi$ in this gauge in (\ref{Thetaint}), we find that it
indeed reproduces the expressions for $\Theta$ that
we obtained in section 3 from the
thermodynamic calculations.

   The construction of the Killing potential (\ref{omegaxi}) by means of
Killing-Yano tensors that we have described  is
essentially unique.  It is interesting, therefore, 
to observe that if we choose not to add
the ``gauge correction'' term $\nu$ to the Killing potential given in
(\ref{omegaxi}), then the integral (\ref{Thetaint}) over the horizon
produces precisely the modified quantity $\Theta'$ that we discussed in 
section 3, which can be written in terms of the geometric 
volume $V'=r_+\, A/(D-1)$
as in (\ref{thetapo}).  It is not clear whether there is some simple
geometrical explanation for this. 
   
\section{Conclusions}

  In this paper, we have investigated some of the consequences of 
treating the cosmological constant, or the gauge coupling constant in a 
gauged supergravity, to become a dynamical variable.  In particular, this means
that it should then be treated as a thermodynamic variable in the
first law of thermodynamics for black holes.  Since the cosmological
constant can be thought of as a pressure, this means that its conjugate
variable in the first law is proportional to a volume.  Using the
first law, we have calculated
this ``thermodynamic volume'' $V$ for a wide variety of black holes, including
static multi-charge solutions in four, five and seven dimensional
gauged supergravities; rotating Kerr-AdS black holes in arbitrary dimensions;
and certain charged rotating black holes in four and five dimensional
gauged supergravities. 

   When there is no rotation, the thermodynamic volume $V$ can be interpreted
as an integral of the scalar potential over the volume ``inside the event
horizon'' of the black hole.  In cases without scalar fields, this 
corresponds precisely to a naive geometrical notion of the ``volume''
inside the horizon.  When there is rotation, however, the thermodynamic
volume $V$ differs from the notion of the ``geometric volume'' $V'$ by
a shift related to the angular momenta of the black hole.  We showed that
although in some examples 
the geometric volume has certain intriguing characteristics
suggestive of a volume in Euclidean space that is ``excluded'' by the
black hole, it appears that the thermodynamic volume has a more
universal character.  In particular, we have found that it and the
horizon area obey the ``Reverse Isoperimetric Inequality'' (\ref{iso1}),
which can be restated as the property that at fixed geometric volume $V$,
the black hole with the largest entropy is Schwarzschild-AdS.

  Although the concept of the thermodynamic volume $V$ requires that one
consider an asymptotically AdS black hole in a theory with a 
nonvanishing cosmological constant, interestingly it is nevertheless 
possible (except in $D=7$) to
take a smooth limit in the expression for $V$ in which the cosmological
constant is set to zero.  Since the thermodynamic volume still, in
general, differs from the geometric volume in this limit, it appears that
to give a definition of $V$ for an asymptotically flat black hole,
one needs first to obtain the expression in the more
general asymptotically AdS case.  For example, for 
the Myers-Perry asymptotically flat
rotating black holes, the thermodynamic volume is given by setting $g=0$
in (\ref{Vkads}).  As we discussed in section 4, this limiting procedure
also works for the known rotating black holes in all gauged supergravities
except in $D=7$.  The case 
$D=7$ is exceptional because of the $g$ dependence of the odd-dimensional
self-duality constraint in the seven-dimensional gauged supergravity.  As
a consequence, the volume diverges in the $g\rightarrow0$ limit if
the electric charges and all three rotation parameters are non-zero.

   We also studied the derivation of the Smarr relation when the cosmological
constant is allowed to become a thermodynamic variable.  This
procedure, which is a generalisation of the Komar method for
asymptotically flat black holes, involves the introduction of a Killing
potential 2-form $\omega$ whose divergence gives the asymptotic
timelike Killing vector.  Because of the gauge freedom to add a co-closed
2-form to $\omega$, the procedure does not provide an unambiguous
computation of the conjugate variable $\Theta$ unless one first fixes the
gauge ambiguity by requiring that the integration at infinity yield
the correct expression for the mass of the black hole. Having made this
gauge choice, we showed that one then recovers the thermodynamic
result for $\Theta$. 

We also presented a method for constructing the
Killing potentials for the Killing vectors in the Kerr-AdS black holes, based
on the existence of conformal Killing-Yano tensors in these metrics.
They occur because of certain ``hidden symmetries'' in the Kerr-AdS metrics,
associated with the separability of equations such as the Dirac 
equation in these backgrounds.  The procedure for constructing the 
Killing potential from the Killing-Yano tensors is an essentially unique
one, and it yields the result in a very specific gauge. Interestingly, 
it is the gauge in which the integral $\int_H {*\omega}$ generates the
``geometric volume'' $V'$.  This suggests that the other remarkable
properties of the geometric volume, such as the fact that it is given by
the Euclidean space formula (\ref{volarea}), might be related to the
existence of the hidden symmetries of the Kerr-AdS metrics.

\section*{Acknowledgments}

M.C. and D.K. are grateful for hospitality at the Mitchell Institute for
Fundamental Physics and Astronomy, during the course of this work.
The research of M.C. is supported in part by DOE grant
DE-FG02-95ER40893 and the  Fay R. and Eugene L. Langberg Chair.
D.K. is supported by the Herchel Smith Postdoctoral Fellowship at the
University of Cambridge.  The research of C.N.P. is
supported in part by DOE grant DE-FG03-95ER40917.

\appendix

\section{Killing Potentials in $D=4$ and $D=5$ Kerr-AdS}

   In this appendix, for illustrative purposes,
we present explicit results for the Killing potentials in the four-dimensional
and five-dimensional Kerr-AdS metrics.

\subsection{$D=4$ Kerr-AdS}

   In the frame that is non-rotating at infinity, the four-dimensional
Kerr-AdS metric, satisfying $R_{\mu\nu}=-3g^2\, g_{\mu\nu}$, can be written
as
\bea
ds_4^2 &=& -\fft{(1+g^2 r^2)\, \Delta_\theta\, dt^2}{\Xi} +
\fft{(r^2+a^2)\sin^2\theta\, d\phi^2}{\Xi} + \fft{\rho^2\, dr^2}{\Delta_r}
+\fft{\rho^2\, d\theta^2}{\Delta_\theta} \nn\\
&& +\fft{2m r}{\Xi^2\, \rho^2}\, 
  (\Delta_\theta\, dt - a \sin^2\theta\, d\phi)^2\,,
\eea
where
\bea
\Delta_r &=& (r^2+a^2)(1+g^2 r^2) -2mr\,,\qquad 
   \Delta_\theta = 1-a^2 g^2 \cos^2\theta\,,\nn\\
\rho^2 &=& r^2 + a^2 \cos^2\theta\,,\qquad \Xi=1 -a^2 g^2\,.
\eea

  The 1-form potential $b$ given by (\ref{b_MPlambda}) is
\be
b= \ft12(r^2 + a^2\sin^2\theta)\, dt -\fft{a\, (r^2+a^2)\sin^2\theta}{2\Xi}\,
(d\phi - a g^2\, dt)\,.
\ee
Following the steps described in section \ref{PCKYsec} for constructing the
Killing potentials $\omega^{(0)}$ and $\omega^{(1)}$ in
this case, we find that their contravariant components are given by
\bea
\omega^{(0)\, tr} &=& -\fft{r(r^2+a^2)}{3\rho^2}\,,\qquad
\omega^{(0)\, t\theta}= -\fft{a^2\sin\theta\cos\theta}{3\rho^2}\,,\nn\\
\omega^{(0)\, r\phi}&=& \fft{ar(1+g^2 r^2)}{3\rho^2}\,,\qquad
\omega^{(0)\, \theta\phi}=\fft{a\Delta_\theta\, \cot\theta}{3\rho^2}\,,\\
\omega^{(1)\, tr} &=& -\fft{a^2 r(r^2+a^2)\cos^2\theta}{\rho^2}\,,\qquad
\omega^{(1)\, t\theta}= \fft{a^2 r^2\sin\theta\cos\theta}{\rho^2}\,,\nn\\
\omega^{(1)\, r\phi}&=& \fft{a^3r(1+g^2 r^2)\cos^2\theta}{\rho^2}\,,\qquad
\omega^{(1)\, \theta\phi}=-\fft{a r^2\Delta_\theta\, \cot\theta}{\rho^2}\,.
\eea
These Killing potentials give rise to the corresponding Killing vectors
\bea
\nabla_\mu \omega^{(0)\mu\nu}\, \del_\nu &=&\fft{\del}{\del t} + a g^2\, 
\fft{\del}{\del\phi}\,,\nn\\
\nabla_\mu \omega^{(1)\mu\nu}\, \del_\nu &=& a^2 \,\fft{\del}{\del t} +
   a\, \fft{\del}{\del\phi}\,.
\eea
Thus the Killing potential for the Killing vector
\be
\xi= \fft{\del}{\del t} +
 \Omega\, \fft{\del}{\del\phi}
\ee
that is null on the horizon is
\be
\omega_\xi = \fft{r_+^2}{(r_+^2+a^2)}\,
\Big( \omega^{(0)} + \fft1{r_+^2}\, \omega^{(1)}\Big)\,.
\ee

\subsection{$D=5$ Kerr-AdS}

  In the frame that is non-rotating at infinity, the five-dimensional
Kerr-AdS metric, satisfying $R_{\mu\nu}=-4g^2\, g_{\mu\nu}$, can be written
as
\bea
ds_5^2 &=&-\fft{(1+g^2 r^2)\,\Delta_\theta\, dt^2}{\Xi_1\Xi_2} +
\fft{(r^2+a_1^2)\sin^2\theta\, d\phi_1^2}{\Xi_1} +
\fft{(r^2+a_2^2)\cos^2\theta\, d\phi_2^2}{\Xi_2} + 
\fft{\rho^2\, dr^2}{\Delta_r} + \fft{\rho^2\, d\theta^2}{\Delta_\theta} \nn\\
&&+ \fft{2m}{\rho^2}\, \Big[ \fft{\Delta_\theta\, dt}{\Xi_1\Xi_2} - 
   \fft{a_1\sin^2\theta\, d\phi_1}{\Xi_1} -
         \fft{a_2\cos^2\theta\, d\phi_2}{\Xi_2}\Big]^2\,,
\eea
where
\bea
\Delta_r &=& \fft{(r^2+a_1^2)(r^2+a_2^2)(1+g^2 r^2)}{r^2}-2m\,,\qquad
 \Delta_\theta= 1- a_1^2 g^2 \cos^2\theta - a_2^2 g^2 \sin^2\theta\,,\nn\\
\rho^2 &=& r^2 + a_1^2\cos^2\theta+ a_2^2\sin^2\theta\,,\qquad
\Xi_1=1-a_1^2 g^2\,,\qquad \Xi_2=1-a_2^2 g^2\,.
\eea

  The 1-form potential $b$ given by (\ref{b_MPlambda}) is
\be
b= \ft12 (r^2+a_1^2\sin^2\theta+a_2^2\cos^2\theta)\, dt -
\fft{a_1(r^2+a_1^2)\sin^2\theta}{2\Xi_1} (d\phi_1-a_1 g^2 dt) -
\fft{a_2(r^2+a_2^2)\cos^2\theta}{2\Xi_2} (d\phi_2-a_2 g^2 dt)\,.
\ee
Following the steps described in section \ref{PCKYsec} for constructing the
Killing potentials $\omega^{(0)}$, $\omega^{(1)}$ and $\omega^{(2)}$ in
this case, we find that their contravariant components are given by
\bea
\omega^{(a)\, tr} &=& -\fft{(r^2+a_1^2)(r^2+a_2i^2))}{4 r\rho^2}\, 
  \Big[1, \fft{2(1-\Delta_\theta)}{g^2}, a_1^2\, a_2^2 \Big]\,,\nn\\
\omega^{(a)\, t\theta} &=&\fft{(a_1^2-a_2^2)\sin\theta\cos\theta}{4\rho^2}\,
\Big[-1, 2r^2,a^2 b^2\Big]\,,\nn\\
\omega^{(a)\, r\phi_1} &=&\fft{a_1(r^2+a_2^2)}{4 r\rho^2}\,
\Big[(1+g^2 r^2), \fft{2(1+g^2 r^2)(1-\Delta_\theta)}{g^2}, a_2^2(r^2+a_1^2)
\Big]\,,\nn\\
\omega^{(a)\, r\phi_2} &=&\fft{a_2(r^2+a_1^2)}{4 r\rho^2}\,
\Big[(1+g^2 r^2), \fft{2(1+g^2 r^2)(1-\Delta_\theta)}{g^2}, a_1^2(r^2+a_2^2)
\Big]\,,\nn\\
\omega^{(a)\, \theta\phi_1} &=&\fft{a_1\cot\theta}{4\rho^2}\, \Big[
\Delta_\theta, -2 r^2\Delta_\theta,a_2^2(a_1^2-a_2^2)\sin^2\theta\Big]\,,\nn\\
\omega^{(a)\, \theta\phi_2} &=&-\fft{a_2\tan\theta}{4\rho^2}\, \Big[
\Delta_\theta, -2 r^2\Delta_\theta,-a_1^2(a_1^2-a_2^2)\cos^2\theta\Big]\,,
\eea
where the components for $\omega^{(a)}$ with $a=0$, 1 and 2 correspond to
the first, second and third entries of the square bracketed factors 
respectively.
 
  The three Killing potentials give rise to the following Killing vectors:
\bea
\nabla_\mu \omega^{(0)\mu\nu}\, \del_\nu &=&\fft{\del}{\del t} + 
  a_1 g^2\, \fft{\del}{\del\phi_1} + a_2 g^2\, \fft{\del}{\del\phi_2}\,,\nn\\
\nabla_\mu \omega^{(1)\mu\nu}\, \del_\nu &=&(a_1^2+a_2^2)\, 
   \fft{\del}{\del t} +
  a_1(1+a_2^2 g^2)\, \fft{\del}{\del\phi_1} +
 a_2(1+a_1^2 g^2)\, \fft{\del}{\del\phi_2}\,,\nn\\
\nabla_\mu \omega^{(2)\mu\nu}\, \del_\nu &=&a_1^2 a_2^2\Big(
\fft{\del}{\del t} +
 \fft{1}{a_1}\, \fft{\del}{\del\phi_1} + 
  \fft{1}{a_2}\, \fft{\del}{\del\phi_2}\Big)\,.
\eea
Thus the Killing potential for the Killing vector
\be
\xi= \fft{\del}{\del t} +
 \Omega_1\, \fft{\del}{\del\phi_1} + \Omega_2\, \fft{\del}{\del\phi_2}
\ee
that is null on the horizon is
\be
\omega_\xi = \fft{r_+^4}{(r_+^2+a_1^2)(r_+^2+a_2^2)}\, 
\Big( \omega^{(0)} + \fft1{r_+^2}\, \omega^{(1)} +
   \fft1{r_+^4}\, \omega^{(2)}\Big)\,.
\ee


\end{document}